
\magnification \magstep 1
\hsize 6 true in
\vsize 8.5 true in
\input mssymb
\openup 1 \jot
\def\today{\ifcase\month\or
  January\or February\or March\or April\or May\or June\or
  July\or August\or September\or October\or November\or December\fi

  \space\number\day, \number\year}

\magnification \magstep 1
\hsize 6 true in
\vsize 8.5 true in
\input mssymb
\nopagenumbers
\openup 1 \jot
\centerline {\bf A JUDGEMENT ON SINORS }
\vskip 1.5cm
\centerline {A CHAMBLIN}
\centerline {\&}
\centerline {G W GIBBONS}
\centerline {D.A.M.T.P.}
\centerline {University of Cambridge}
\centerline {Silver Street}
\centerline {Cambridge CB3 9EW}
\centerline {U.K.}

\vskip 1.5cm
\centerline {\rm \today}
\vskip 1.5cm
\centerline  {\bf ABSTRACT}
\tenrm
{\narrower \narrower \smallskip
This note contains some  comments on
a recent paper by Friedman [1]
on two-component spinors in spacetimes which do not admit a time-orientation,
and is intended to clarify the relation of the work reported in that paper
to previous literature.}

\beginsection Notation

In [1] Friedman introduces two two-fold covers of the group
$L_+$
of space-orientation preserving Lorentz transformations,
 the so-called \lq ortho-chirous Lorentz group \rq. In fact there are eight
covers
of the full Lorentz group, called by Dabrowski [2] ${\rm Pin} ^{a.b.c}$. If
$\cal  P$ and $\cal  T$ cover $P$ and $T$ respectively then $a,b,c$,  may be
determined (in four spacetime dimensions and using his conventions) from the
relations:
$$
{\cal P}^2 = -a
\eqno (1)
$$
$${\cal
T }^2 = b
\eqno(2)
$$
and
$$
{\cal P  T} = abc {\cal T  P}.
\eqno(3)
$$
The   Cliffordian groups
$${\rm Pin} ^{+ - + }
\subset{\rm Cliff}(3,1;{\Bbb R})\equiv {\Bbb R} (4)\eqno(4)$$
and
$${\rm Pin }^{- + +} \subset {\rm Cliff}(1,3; {\Bbb R}) \equiv {\Bbb
H}(2)\eqno(5)$$ associated with signature $+ + + -$ and $- - - +$
respectively are obtained by representing reflections by  Clifford
multiplication, i.e. one has ${\cal P} = \gamma _1 \gamma _2 \gamma _3$ and
${\cal T}=\gamma _0$. The Cliffordian groups act on spinors by Clifford
multiplication. We shall call this the   Cliffordian action of ${\rm Pin} ^{+ -
 + }$ or
${\rm Pin }^{-  +  +}$ on spinors respectively. In the Cliffordian case,
the discrete groups generated by ${\cal P}$ and ${\cal T}$ are subgroups
of the discrete groups ${\rm Dirac}(3,1)$ or ${\rm Dirac}( 1,3)$ which
double cover the groups generated by all possible reflections in four
orthogonal directions.

We note {\it en passant} that the reason for the minus sign in (1) is that $a$
was originally  defined
(in all spacetime dimensions) by the sign
of the cover of a {\sl single} spatial reflection.
A consequence of (3) is
$$
({\cal P T}) ^ 2 = -c.
\eqno(6)
$$
Thus  Friedman's Sinor groups, since he is not concerned with the
action of space-reflections,  are the images of four  2-1 forgetful
homomorphisms:
$$
 {\rm Pin}  ^{{\pm} + {\pm} } \rightarrow {\rm Sin} ^+
\eqno(7)
$$
and
$$
 {\rm Pin}^ {\pm - {\pm} } \rightarrow {\rm Sin} ^-.
\eqno(8)
$$
We propose calling quantities transforming under the action of ${\rm Sin}^\pm$
\lq sinors \rq .

\beginsection Weyl Sinors

It is well-known that one cannot represent time-reversal in a complex-linear
fashion on two-component Weyl spinors. Although
one may retain the complex notation one is in effect
working with a real four dimensional vector bundle whose structural group
is ${\rm Sin}^\pm$.
Anti-linear
actions of both $\cal T$ and $\cal P$  on two
component spinors has been considered previously by Staruskiewicz [3]. From the
point of view of the Penrose sphere
construction, i.e. thinking projectively,
The actions of both of $ T$
and $ P$ corresponds to the anti-podal map on the the two-sphere $\equiv
{\Bbb C}{\Bbb P}^1$. The possible  spinors differ in how
the action of $P$ and $T$  is lifted to ${\Bbb C}^2$.

 Staruskiewicz considers what he calls spinors of two \lq  kinds \lq. From his
table III it follows that
for  both kinds ${\cal T}^2=-1$ . Thus as far as time-reversal is concerned it
is
Friedman's ${\rm Sin}^-$ which is involved. Staruskiewicz
takes for the action
of $\cal P$ either $i{\cal T}$ for the first kind or $\cal T$
for the third kind. It follows that $({\cal PT})^2=+1$ for the first
kind while $({\cal PT}) ^2 = -1$ for the third kind.  In Dabrowski's
notation the first kind  corresponds to ${\rm Pin} ^{ - - - }$
and the third
kind to ${\rm Pin}^ {+  -  + }$.
Thus the group involved for spinors of the third kind is {\rm Cliff}ordian
(with respect to signature $+++- )$.
Note that what he calls space-reflections, i.e. $\cal P$  might
well, from a physical point of view, be called $\cal CP$ , where $\cal C$ is
charge conjugation
if one were thinking of its action on the solutions of the two-component Weyl
neutrino equations, rather than just a spinor at one point in spacetime.

\beginsection Majorana Sinors

In order to relate the Staruskiewicz' and Friedman's Weyl spinor formalism to
the {\rm Cliff}ord algebra approach, we consider,
as an  alternative to using complex two component spinors, the use
of four real component Majorana spinors. Since ${\rm Cliff}(3,1;{\Bbb R})
\equiv {\Bbb R}(4)$, the algebra of four by four real matrices,  this is most
conveniently
done using signature $+ + + -$. In other words we adopt $\gamma _0 ^2 = -1$
and $\gamma _i ^2=1$. The relationship to the two-component formalism is that
the {\rm Cliff}ordian cover of total reflection $\cal PT$, i.e.
$\gamma_5 = \gamma _1 \gamma_2 \gamma _3 \gamma _0$ acting on ${\Bbb R}^4_{\rm
Majorana} \equiv {\Bbb C}^2 _{\rm Weyl}$ serves as a complex structure [4,5].
Projectively speaking Majorana spinors correspond to points in ${\Bbb R}{\Bbb
P}^3$
and the Dirac group ${\rm Dirac}(3,1) /{\pm}1$ acts as the symmetry group of
the Kummer configuration [5].

The {\rm Cliff}ordian choice ${\rm Pin} ^{+ - +}$ is:
$$
{\cal P}= \gamma _1 \gamma _2 \gamma _3
\eqno(9)$$
and
$$
{\cal T}=\gamma _0
\eqno(10)
$$
Since $\gamma_5$ anti-commutes with $\gamma _1 \gamma _2 \gamma _3$ and
$\gamma_0$ both ${\cal P}$ and $\cal T$ are anti-linear considered as acting on
${\Bbb C}^2_{\rm Weyl}$ and thus we see  the Cliffordian
action corresponds precisely to Staruskiewicz's spinors of the third kind.
His spinors of the first kind correspond to the choice
$$
{\cal P} = \gamma _0
\eqno(11)
$$
and
$$
{\cal T} = \gamma _0
\eqno(12)
$$
which gives an action of the non-{\rm Cliff}ordian group ${\rm Pin} ^{+ - +}$.

This relationship between real Majorana spinors and complex  Weyl spinors has
an analogue for Maxwell's equations. One may think of the Hodge star
operator, which acting on 2-forms on an orientable spacetime of signature $+ +
+ -$ or $- - - +$
satisfies $\star ^2 = -1$, as a complex structure on the space of Maxwell
fields. This is behind the well-known complex notation in which one writes
Maxwell's equations in the first-order form as:
$$
\nabla \times ({\bf E} +i{\bf B}) = i {{\partial }\over {\partial t}} ( {\bf E}
+ i {\bf B})
$$
or in a more covariant notation

$$
( d + \delta) F=0.
$$
where $\delta$ is the adjoint of $d$.

Physically speaking the space of complex self-dual Maxwell fields
describes photons of a particular helicity, for example  circularly polarized
photons with a right-handed polarization .
On a non-orientable  spacetime this reformulation of
Maxwell's equations is not possible. In other words
on a non-orientable spacetime there is an obstruction to introducing
this particular complex structure on the space of Maxwell fields. The physical
interpretation is that on such spacetime   it is not possible to speak
of a right or left-handed photon.
For some other speculations concerning the non-orientable case see [8].

\beginsection Cliffordian versus non-Cliffordian

Evidently there are many possibilities for representing $P$ and $T$.
The fact that some of the groups are non-Cliffordian or if they are
Cliffordian they have non-Cliffordian  actions
is not in itself of great physical significance. Consider
 the classical Dirac equation in flat space with a mass:
$$
\gamma ^\mu \partial _\mu \psi + m \psi =0.
\eqno(13)
$$
This is invariant under time reversal in the sense that:
$$
\psi (t, {\bf x}) \rightarrow \gamma_1 \gamma _2 \gamma _3 \psi (-t, {\bf x})
\eqno(14)
$$
takes solutions to solutions, but $\cal T$ is {\sl not}
represented at even one point of spacetime by $\gamma _0$, rather since
$$
\psi (t, {\bf x}) \rightarrow \gamma_0 \psi (t, -{\bf x}),
\eqno(15)
$$
also takes soluions to solutions  the {\sl group } is the Cliffordian group
${\rm Pin}^{+ - +}$ but its action is not Cliffordian.

On the other hand, the standard actions of $\cal P $ and $\cal T$
on the full second quantized Hilbert space
$\cal H$ in flat spacetime quantum field theory is  of a non-Cliffordian
group.
Conventionally one seeks actions which take positive energy states to
positive energy states. To that end one follows Wigner and chooses
$\cal P$ to
act unitarily and scales it such that ${\cal P}^2={\pm}1$. By contrast,
again following Wigner, one chooses $\cal T$ to act anti-unitarily
(on fermion  states) and finds that $ {\cal T}^2 = -1$.
Since in addition $\cal P$ and $\cal T$ anti-commute
the relevant groups are  ${\rm Pin}^{\pm1 -1 -1}$. These include the group
associated to Staruskiewiz's spinors of the first kind,
though of course it acts on a different space. In fact in flat space
quantum field theory the Hilbert space $\cal H$ is constructed from
the space of solutions of the Dirac equation and these carry other
possible actions of $\cal P$, $\cal T$ and $\cal PT$,
both linear and anti-linear, in addition to the action of
\lq charge conjugation\rq $\cal C$. Thus charge conjugation acts by
\lq complex conjugation\rq  on the classical solutions but acting on
$\cal H$ it acts unitarily, taking particle to anti-particle states.
The resolution of this apparent paradox is that the complex structure
on $\cal H$
should be distinguished from that used on the space of classical
solutions [5].

\beginsection Obstructions to Sinors

In flat spacetime the distinctions we have drawn above between the various Pin
groups and their actions might seem to have little physical consequence.
However spacetime is not flat and may well not be space and time orientable.
The  distinctions then become vital because the obstructions to the
global existence of a given  Pinor bundle
depends upon precisely what Pin group we are considering.

Karoubi has shown [9]that the obstruction to a {\sl Cliffordian} ${\rm Sin}^
{\pm}$
structure is
$$
w_2(M) + w_1^-(M, g_L) \smile w_1^-(M, g_L).
$$
where $w_2(M)$ is the second Stiefel-Whitney class of $TM$ the tangent bundle
of $M$ and $w_1^-(M, g_L)$ is the first Stiefel-Whitney class of the bundle
associated
to the negative sign in the metric. Note that $w_2(M)$ is a topological
invariant independent of the existence of any metric on $M$.

Thus for signature $+++-$, i.e. for ${\rm Pin}^{+-+} \sim {\rm Sin}^-$ we have
$$
w_1^-(M,g_L)=w_1^T(M,g_L)
$$
where $w_1^T(M,g_L )$ is the element of $H^1(M;{\Bbb Z}_2)$ giving the
obstruction to time-orientability.
On the other hand for signature $---+$, i.e for ${\rm Pin} ^{-++} \sim{\rm
Sin}^+$
$$
w_1^-(M,g_L)= w_1^S(M,g_L)
$$
where $w_1^S(M,g_l)$ is the obstruction to space-orientability.
Note that both $w_1^T(M, g_L)$ and $w_1^S(M,g_l)$ depend on the existence of a
Lorentzian
metric $g_L$ but their sum is a topological invariant:
$$
w_1^S(M, g_L) + w_1^T(M.g_L )= w_1(M)
$$
where $w_1(M)$ is the first Stiefel-Whitney class of the tangent bundle $TM$
and is the obstruction to time orientability. If $w_2^{\pm}$ are the
second Stiefel-Whitney classes of the indicated bundles then
$$
w_2(M)= w_2^+(M,g_L)+w_2^-(M,g_L)+ w_1^+(M,g_L) \smile w_1^-(M,g_L).
$$
However the second Stiefel-Whitney class of a one-dimensional bundle vanishes
and thus
$$
w_2(M)=w_2^S(M,g_L) + w_1^T(M,g_L) \smile w_1^S(M,g_L).
$$

Thus the ${\rm Pin}^{+-+} \sim {\rm Sin}^-$ the obstruction is
$$
w_2(M) + w_1^T(M,g_L) \smile w_1^T(M,g_L),
$$
while for ${\rm Pin}^{-++} \sim {\rm Sin}^+$ the obstruction
is
$$
w_2(M) + w_1^S(M,g_L) \smile w_1^S(M,g_L).
$$

Consider, as does Friedman, a space-orientable but time non-orientable
spacetime. Then $w_1^S(M,g_L)=0$. The obstruction to ${\rm Pin}^{+-+} \sim {\rm
Sin}^-$ is therefore
$$
w_2(M) +w_1^T(M,g_l) \smile w_1^T(M,g_L)
$$
while to ${\rm Pin}^{-++} \sim {\rm Sin}^+$ it is just
$$
w_2(M).
$$

A particularly interesting case is
$\{ M,g_L \} \equiv$  antipodally identified De-Sitter spacetime.
Topologically we may think of $ M$  as ${\Bbb R}{\Bbb P}^4 -\{{\rm pt.}\}$.
This has the same first and second Stiefel-Whitney classes as ${\Bbb R} {\Bbb P
}^4$, thus $w_2(M)=0$ . Moreover we clearly have for this choice of metric that
$
w_1^T(M,g_L)$ is given by the unique element of $H^1(M)$. It follws immediately
that antipodally identified De-Sitter spacetime admits a ${\rm Pin}^{-++} \sim
{\rm Sin}^+$ structure but not a ${\rm Pin}^{+-+} \sim {\rm Sin} ^-$
  structure. This is in agreement with a direct construction of the relevant
bundles carried out by one of us with
L Dabrowski some time ago and reported in [10] It also agrees with Friedman.

In fact the argument will generalize to manifolds of the form $ M=\tilde \Sigma
\times {\Bbb R} /A$ where $A=T \circ I$ where $T$ is time-reversal on the $\Bbb
R$ factor, which corresponds to time, and $I$ is an space-orientation
preserving involution which acts
freely on $\tilde \Sigma$. We claim that $w_2(M)=0$. To see this we use
proposition 4 of [11] which states that if the tangent bundle of a 4-manifold
admits
three linearly independent non-vanishing cross-sections then $w_k(M)=0$
for $k>1$. To construct the requisite cross-sections we
parallelize $\Sigma /I$ and lift the parallelization to $\Sigma$.
The three cross-sections are invariant under the action of $T$ and thus descend
to $M$. Thus, in accord with Friedman's result, all manifolds of this
type admit ${ \rm Sin }^+$, and admit ${ \rm Sin }^-$ if and only if
$$
w_ 1^T(M) \smile w_1^T(M) =0.
$$

\beginsection References

\medskip
\noindent[1] J. L. Friedman {\it Two-component spinor fields in a class of
time-nonorientable spacetimes } WISC-MILW-94-TH-24
\medskip
\noindent [2] L. Dabrowski {\it Group Actions on Spinors } Bibliopolis, Napoli
(1988)
\medskip
\noindent [3] A. Staruskiewicz {\it On  two kinds of spinors}  Acta Physica
Polonica {\bf 7B}
557-565 (1976)
\medskip
\noindent [4] G. W. Gibbons \& H-J Pohle Complex Numbers, Quantum Mechanics and
the Beginning of {\sl Nucl. Phys.} {\bf B 410} 117-142 (1993)
\medskip
\noindent [5] G. W. Gibbons The Kummer Configuration and the Geometry of
Majorana Spinors
in {\sl Spinors, Twistors, Clifford Algebras and Quantum Deformations}
eds. Z. Oziewicz et al., Kluwer, Amsterdam (1993)
\medskip
\noindent [6] A. Chamblin, {\it On the Obstructions to Non-{\rm Cliff}ordian
Pin
Structures}, Comm. Math. Phys. {\bf 164}, No. 1, pgs. 65--87 (1994).
\medskip
\noindent [7] S. Carlip and C. DeWitt-Morette, {\it Where the Sign of the
Metric Makes
a Difference}, Phys. Rev. Lett. {\bf 60}, pg. 1599 (1988)
\medskip
\noindent [8] R. Sorkin {\it J Phys } {\bf A10} 717-725 (1977)
\medskip
\noindent [9] M. Karoubi {\sl Ann Scient \'Ecole  Normale  } {\bf 1} 161 (1968)
\medskip
\noindent [10] G. Gibbons {\it Int. Journ. Mod. Phys.} {\bf 3} 61 (1994)
\medskip
\noindent [11] J. Milnor and J. Stasheff {\it Characteristic Classes} Annals of
Mathematics Studies 76 Princeton University Press (1974)

\bye